\documentclass[journal]{IEEEtran}
\usepackage{cite}

\ifCLASSINFOpdf
  \usepackage[pdftex]{graphicx}
  \graphicspath{{./figs/}}
\else
  \usepackage[dvips]{graphicx}
\fi
\usepackage{booktabs}

\usepackage{amsmath}
\usepackage{amssymb}

\usepackage{multirow}

\ifCLASSOPTIONcompsoc
  \usepackage[caption=false,font=normalsize,labelfont=sf,textfont=sf]{subfig}
\else
  \usepackage[caption=false,font=footnotesize]{subfig}
\fi
\usepackage{dblfloatfix}
\begin{document}
%
\title{Generalized Matrix-Pencil Approach to Estimation of Complex Exponentials with Gapped Data}
%
%
%

\author{Jianping~Wang 
        and~Alexander~Yarovoy
\thanks{The authors are with the Faculty
of Electrical Engineering, Mathematics and Computer Science (EEMCS), Delft University of Technology, Delft, 2628CD the Netherlands e-mail: J.Wang-4@tudelft.nl, A.Yarovoy@tudelft.nl.}
}

\maketitle

\begin{abstract}
A generalized matrix-pencil approach is proposed for the estimation of complex exponential components with segmented signal samples, which is very efficient and provides super-resolution estimations. It is applicable to the signals sampled segmentally with the same sampling frequency and direction of arrival (DOA) estimation with distributed arrays within which array elements are placed uniformly with the same inter-element spacing.  
\end{abstract}

\begin{IEEEkeywords}
Generalized Matrix Pencil Approach, Super-resolution Estimation, Complex Exponential, Segmented Samples, Signal Estimation.
\end{IEEEkeywords}

%
\IEEEpeerreviewmaketitle

\section{Introduction}
%
%
%
%


Harmonic retrival problem is a general problem of estimating the frequencies and damping factors from the measurements taken in time, space, etc. 
It is widely used in the field of radar, sonar, communication, radio astronomy, and so on.

In the literature, the estimation approaches of  undamped/damped exponential components which are generally developed based on parametric models. The existing approaches are typically rely on the analysis of state-space of measurements and are implemented via space-searching technique or search-free strategies. The classical space-searching approaches include MUltiple SIgnal Classification (MUSIC), the Amplitude and Phase EStimation method (APES) \cite{stoica2005spectral} and the Iterative Adaptive Approach (IAA) \cite{Yardibi2010TAES} while typical search-free approaches contain the Estimation of Signal Parameters via Rational Invariance Techniques (ESPRIT) algorithm \cite{Roy1986TASSP_ESPRIT} and matrix pencil-type approaches \cite{Sarkar1995}. All these approaches have been extended to multidimensional cases for various applications.

All the above approaches are developed for harmonic retrieval with continuously sampled data with a fixed sampling interval. However, in practice, due to sensor failure, interference or noise effect, memory constraint, etc, gapped signals are frequently acquired as a few separate segments. In such cases, the aforementioned estimation methods are not applicable due to the break of certain structures implicitly used. To tackle this problem, several methods have been introduced for harmonic retrieval with gapped data \cite{Zhou1992SECON,Lee2009ICASSP,Stoica2000Astron,WANG2005DSP,Stoica2009SPL,Broersen2006DSP,Babu2010DSP,Knight2012SC,Vu2012JSTSP,Parvazi2012TSP,Bian2016DSP,Baghi2016PRD,Zou2016SR,Wang2018TGRS}. In \cite{Stoica2000Astron}, the generalized APES (GAPES) method was introduced to deal with the gapped data based on Fourier basis grid search. For more general missing data pattern, missing-data APES (MAPES) \cite{WANG2005DSP} was developed using the  expectation maximization (EM) algorithm \cite{WANG2005DSP}. Meanwhile, the missing-data IAA (MIAA)~\cite{Stoica2009SPL} was introduced based on the IAA to estimate the harmonic frequencies and the missing samples. Both MAPES and MIAA are search-based methods but MIAA, as the author claimed, is much faster than the MAPES. However, both methods have difficulties in tackle damping harmonics estimation. On the other hand, the weighted multiple invariance (WMI) approach \cite{Parvazi2012TSP} was introduced based on a linear combination of the rank-reduction criteria obtained shift-invariances of the signal model. Using the MUSIC criterion, the WMI can in principle get accurate estimation; however, the WMI based on polynomial intersection is computationally very sensitive and not stable. Moreover, similar to the MAPES, a Gaussian regression method \cite{Baghi2016PRD} is introduced for spectral estimation with missing data based on the EM algorithm but it requires the power spectral density sufficiently smooth.  

In this paper, taking advantage of the search-free MPA, we introduce the generalized MPA (GMPA) for estimating harmonic components with gapped data. The proposed GMPA exploits the shift-invariance of the signal space of the harmonic components and utilize the singular value decomposition to figure out the signal poles of the constructed Hankel-like matrix. It is accurate and computational very efficient. Although the propper approach has been directly used for signal fusion in \cite{Zou2016SR,Wang2018TGRS}, rigorous analysis was missing. To fill this gap, a mathematical derivation is provided in this paper.

The remainder of the paper is organized as follows. In section~\ref{sec:sig_model}, the generic signal model for harmonic retrieval with gapped data is presented and section~\ref{sec:GMPA} introduces the proposed generalized matrix pencil approach. After that, some numerical simulations are shown in section~\ref{sec:num_simu} to demonstrate the performance of the proposed GMPA and comparisons with the state-of-the-art methods. Finally, conclusions are drawn in section~\ref{sec:conclusion}.

\section{Signal Model} \label{sec:sig_model}
Signal estimation of damped/undamped complex exponential components is a popular problem in signal processing. The damped/undamped complex exponential signal can be generally expressed as 
\begin{equation} \label{eq:sig_model_continuous}
    s(t) = \sum_{n=1}^N \alpha_n \exp\left[(\beta_n + j\omega_n) t \right]
\end{equation}
where $j=\sqrt{-1}$ is the unit of imaginary number. $\alpha_n$, $\beta_n$ and $\omega_n$ are the amplitude, damping factor and angular frequency of the $n^\text{th}$ complex exponential component, respectively. For the undamped exponential components, their damping factor are zero in \eqref{eq:sig_model_continuous}.

Nowadays, the signals in \eqref{eq:sig_model_continuous} are generally sampled and their discrete samples are processed to estimate the signal poles and their related amplitude coefficients. However, due to hardware limitation, memory constraint, interference/impulse noise suppression, the signal samples would be acquired as a few discontinuous segments. Then, the segmented signal samples can be represented as  
\begin{align} \label{eq:sig_model_discrete}
    y_i[k] &= s_i[k] + n_i[k]  \nonumber\\ 
    &= \sum_{n=1}^N \alpha_n \exp\left[(\beta_n + j \omega_n) (k \Delta t + t_i) \right] + n_i[k] \nonumber\\
    &= \sum_{n=1}^N \Tilde{\alpha}_{in} z_n^k + n_i[k], \qquad k=0,\,1,\,\cdots,\, K_i-1;  \nonumber\\
    &\qquad\qquad\qquad\qquad\qquad\quad\,   i=1,\, 2,\, \cdots,\, M
\end{align}  
where $M$ is the number of segments of discrete signal samples and $K_i$ is the number of samples acquired in the $i^\text{th}$ segment. $\Delta t$ is the common sampling interval of all discrete signal segments, and $\Tilde{\alpha}_{in} = \alpha_n e^{(\beta_n + j \omega_n)t_i}$ and $z_n = e^{(\beta_n + j \omega_n) \Delta t}$. $t_i$ is the starting sampling time of the $i^\text{th}$ signal segment and \textit{the time delays between adjacent signal segments can be arbitrary}. $n_i$ represents the measurement errors and thermal noise of the system.

\section{Generalized Matrix Pencil Approach} \label{sec:GMPA}
\subsection{Review of Matrix Pencil Approach}
For the uniformly sampled sample in a segment, the Matrix-Pencil approach (MPA) is a popular and efficient method for high-resolution estimation of complex exponential components. The principle of the MPA is described as follows. Considering the noiseless signal samples in the $i^\text{th}$ segment, a Hankel matrix can be constructed as 
\begin{equation}
    \mathbf{H}_{i} = \begin{pmatrix}
    s_i[0] & s_i[1] &  \cdots & s_i[K_i-p] \\
    s_i[1] & s_i[2] &  \cdots & s_i[K_i-p+1] \\
    \vdots & \vdots &  \ddots & \vdots \\
    s_i[p-2] & s_i[p-1] &  \cdots & s_i[K_i-2] \\
    s_i[p-1] & s_i[p] &  \cdots & s_i[K_i-1]
    \end{pmatrix}
\end{equation}
where $p$ is a parameter of the matrix pencil with $M<p<K_i-M$. By removing the first row and last row of $\mathbf{H}_i$, respectively, two new Hankel matrices $\mathbf{H}_i^\uparrow$ and $\mathbf{H}_i^\downarrow$ are obtained. As a Hankel matrix generally has a Vandermonde decomposition, then $\mathbf{H}_i^\downarrow$ and $\mathbf{H}_i^\uparrow$ can be written as  
\begin{equation}\label{eq:H_i_downarrow}
    \mathbf{H}_i^\downarrow = \mathbf{V}_L \mathbf{\Sigma}_i^\downarrow \mathbf{V}_{iR}^H 
\end{equation}
\begin{equation}\label{eq:H_i_uparrow}
    \mathbf{H}_i^\uparrow = \mathbf{V}_L \mathbf{\Sigma}_i^\uparrow \mathbf{V}_{iR}^H
\end{equation}
where 
\begin{equation}
    \mathbf{V}_L = \begin{pmatrix}
    1 & 1 & \cdots & 1 \\
    z_1 & z_2 & \cdots & z_N \\
    z_1^2 & z_2^2 & \cdots & z_N^2 \\
    \vdots & \vdots & \ddots & \vdots \\
    z_1^{p-1} & z_2^{p-1} & \cdots & z_N^{p-1}
    \end{pmatrix} 
\end{equation}
and 
\begin{equation}
    \mathbf{V}_{iR} = \begin{pmatrix}
    1 & 1 & \cdots & 1 \\
    z_1 & z_2 & \cdots & z_N \\
    \vdots & \vdots & \ddots & \vdots \\
    z_1^{K_i-p+1} & z_2^{K_i-p+1} & \cdots & z_N^{K_i-p+1}
    \end{pmatrix}    
\end{equation}
are two Vandermonde matrices. $\mathbf{\Sigma}_i^\downarrow$ and $\mathbf{\Sigma}_i^\uparrow$ are two diagonal matrices, which are given by
\begin{align}
    \mathbf{\Sigma}_i^\downarrow &= \textbf{diag}\left(\left[\Tilde{\alpha}_{i1}, \Tilde{\alpha}_{i2}, \cdots,\Tilde{\alpha}_{iN} \right]\right) \\
    \mathbf{\Sigma}_i^\uparrow &= \mathbf{\Sigma}_i^\downarrow \cdot \textbf{diag}\left( \left[z_1, z_2, \cdots, z_N \right]\right) 
\end{align}
where $\textbf{diag}(\mathbf{a})$ constructs a diagonal matrix with the elements of the vector $\mathbf{a}$ as the diagonal entries.

According to \eqref{eq:H_i_downarrow} and \eqref{eq:H_i_uparrow}, the signal poles $z_n$ can be obtained by solving a generalized eigenvalue problem 
\begin{equation} \label{eq:generalized_EVP}
    \textbf{det} \left( \mathbf{H}_i^\uparrow - \lambda \mathbf{H}_i^\downarrow \right)=0
\end{equation}
where $\textbf{det}(\mathbf{\cdot})$ is the determinant operator. The obtained eigenvalues $\lambda$ would be the signal poles $z_n$, $n=1,\, 2,\, \cdots,\, N$.

In the presence of noise, two Hankel matrices $\Bar{\mathbf{H}}_i^\downarrow$ and $\Bar{\mathbf{H}}_i^\uparrow$ can be constructed by replacing the clean signal data $s_i$ in $\mathbf{H}_i^\downarrow$ and $\mathbf{H}_i^\uparrow$ with $y_i$ in \eqref{eq:sig_model_discrete}. Then, by replacing $\mathbf{H}_i^\uparrow$ and $\mathbf{H}_i^\downarrow$ with $\Bar{\mathbf{H}}_i^\uparrow$ and $\Bar{\mathbf{H}}_i^\downarrow$ in \eqref{eq:generalized_EVP}, the signal poles $z_n$ can be estimated by using the total least-squares matrix pencil \cite{Sarkar1995}. After that, the complex amplitude coefficients $\Tilde{\alpha}_{in}$ can be obtained by solving the system of linear equations in \eqref{eq:sig_model_discrete} using the least-square method.

The aforementioned MPA works well for uniformly sampled signals in a continuous segment. However, for segmented signal samples, it is not straightforwardly applicable. To address this problem, we propose the generalized MPA for segmented signal samples below.

\subsection{Generalized MPA for Segmented Signal Samples}
Considering $M$ segments of noiseless signal samples, one can construct two Hankel matrices $\mathbf{H}^\downarrow$ and $\mathbf{H}^\uparrow$ by  horizontally stacking the Hankel matrices $\mathbf{H}_i^\downarrow$ and $\mathbf{H}_i^\uparrow$, which are given by
\begin{align}
    \mathbf{H}^\downarrow &= \left[\mathbf{H}_1^\downarrow,\, \mathbf{H}_2^\downarrow,\, \cdots,\, \mathbf{H}_M^\downarrow \right]  
 \\
    \mathbf{H}^\uparrow &= \left[ \mathbf{H}_1^\uparrow,\, \mathbf{H}_2^\uparrow,\,\cdots,\, \mathbf{H}_M^\uparrow \right]  
\end{align}
As the signals in all segments satisfy the same model in \eqref{eq:sig_model_discrete}, $\mathbf{H}^\downarrow$ and $\mathbf{H}^\uparrow$ can be represented as the following Vandermonde decomposition
\begin{align} \label{eq:H_down_arrow}
    \mathbf{H}^\downarrow &= \mathbf{V}_L^{tb} \mathbf{\Sigma}^\downarrow \mathbf{V}_R^H \\ \label{eq:H_up_arrow}
    \mathbf{H}^\uparrow &=  \mathbf{V}_L^{tb} \mathbf{\Sigma}^\uparrow \mathbf{V}_R^H
\end{align}
with
\begin{equation} \label{eq:V_L_tb}
    \mathbf{V}_L^{tb} = \textbf{repmat}\left(\mathbf{V}_L,[1,M] \right)    
\end{equation}
\begin{equation} \label{eq:sigma_downarrow}
    \mathbf{\Sigma}^\downarrow = \begin{pmatrix}
    \mathbf{\Sigma}_1^\downarrow & \mathbf{0} & \cdots & \mathbf{0} \\
    \mathbf{0} & \mathbf{\Sigma}_2^\downarrow & \cdots & \mathbf{0} \\
    \vdots & \vdots & \ddots & \vdots \\
    \mathbf{0} & \mathbf{0} & \cdots & \mathbf{\Sigma}_M^\downarrow
    \end{pmatrix}    
\end{equation}
\begin{equation} \label{eq:sigma_uparrow}
    \mathbf{\Sigma}^\uparrow = \mathbf{\Sigma}^\downarrow \cdot \textbf{diag}(\textbf{repmat}([z_1,z_2,\cdots,z_N],[1,M]))    
\end{equation}
\begin{equation} \label{eq:V_R}
    \mathbf{V}_R =\begin{pmatrix}
    \mathbf{V}_{1R} & \mathbf{0} & \cdots & \mathbf{0} \\
    \mathbf{0} & \mathbf{V}_{2R} & \cdots & \mathbf{0} \\
    \vdots & \vdots & \ddots & \vdots \\
    \mathbf{0} & \mathbf{0} & \cdots & \mathbf{V}_{MR}
    \end{pmatrix}
\end{equation}
where $\textbf{repmat}(\mathbf{A},[r_1, r_2])$ returns a matrix containing $r_1$ and $r_2$ copies of $\mathbf{A}$ in the column and row dimensions, respectively. 

From \eqref{eq:V_L_tb}, one can see that both the ranks of $\mathbf{V}_L^{tb}$ and $\mathbf{V}_L$ are equal to $N$. So, the ranks of $\mathbf{H}^\downarrow$ and $\mathbf{H}^\uparrow$ are also $N$. According to \eqref{eq:H_down_arrow}-\eqref{eq:V_R}, the rank of $\mathbf{H}^\uparrow - \lambda \mathbf{H}^\downarrow$ is $N-1$ if $\lambda=z_n$, $n=1,\, 2,\, \cdots,\, N$. That is to say, the signal poles $z_n$, $n=1,\, 2,\, \cdots,\, N$ are the eigenvalues of the matrix pencil $\mathbf{H}^\uparrow - \lambda \mathbf{H}^\downarrow$(i.e., the solutions of \eqref{eq:generalized_EVP} when $\mathbf{H}_i^\uparrow$ and $\mathbf{H}_i^\downarrow$ are substituted by $\mathbf{H}^\uparrow$ and $\mathbf{H}^\downarrow$).  

For the noisy signal measurements in practice, the counterparts of the Hankel matrices $\mathbf{H}^\uparrow$ and $\mathbf{H}^\downarrow$ can be formed by replacing $s_i$ with the measurements $y_i$, denoted as $\Bar{\mathbf{H}}^\uparrow$ and $\Bar{\mathbf{H}}^\downarrow$, respectively. Then, following the workflow of the traditional total least-squares matrix pencil, the signal poles and the complex amplitudes can be estimated with multiple segments of signal samples.  

\textit{\textbf{Remark:} The generalized MPA would be simplified to be the traditional MPA when the number of the segments of signal samples reduces to one.}

\section{Numerical Simulation} \label{sec:num_simu}
In this section, some numerical simulations are presented to demonstrate the application of proposed GMPA.

\subsection{Segmented signals with gaps of an integer number of sampling intervals}\label{subsec:seg_signal_integ_TD}

\begin{figure}
    \centering
    \includegraphics[width=0.35\textwidth]{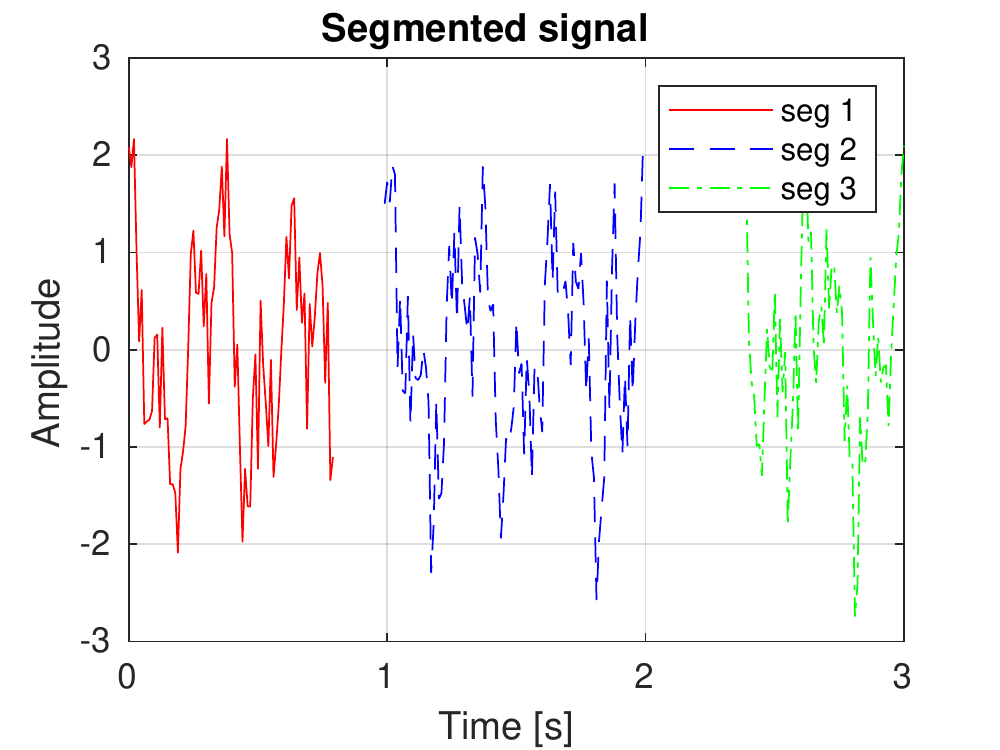}
    \caption{The segmented signal samples of the undamped signal $e^{j6\pi t} + 0.8e^{j16\pi t}$ with the SNR of $5\,\mathrm{dB}$.} 
    \label{fig:Simu_SegSig}
\end{figure}

In this first simulation, we used a two pole signal model $s[n]=e^{j2\pi\cdot 3 n\cdot \Delta t} + 0.8e^{j2\pi \cdot 8 n\cdot \Delta t}$ to generate a signal of three seconds, i.e., 301 samples with $n=0,1,\cdots,300$ and $\Delta t = 0.01\,\mathrm{s}$. The signal samples with $n=80,\, 81,\, \cdots,\,98$ and $n=210,\, 211,\,\cdots,\, 238$ were removed to form three signal segments with samples of 80, 111 and 61, respectively. Moreover, white Gaussian noise was added to synthesize the signal with the SNR of $5\,\mathrm{dB}$ (see Fig.~\ref{fig:Simu_SegSig}).

Due to the non-contiguous sampleing of the three signal segments, we took advantage of the traditional MPA to estimate the signal poles based on the signal in each segment individually, and used the proposed GMPA to jointly process all the data in three segments. The estimated signal frequencies are shown in Table~\ref{tab:Simu_Est_Freq}. One can see that the GMPA with all the three signal segments achieves the smallest RMSE of the estimations of the two frequencies compared to the results obtained with each individual signal segment.  

\begin{figure}[!t]
    \centering
    \vspace{-2mm}
    \subfloat[]{
        \includegraphics[width = 0.36\textwidth]{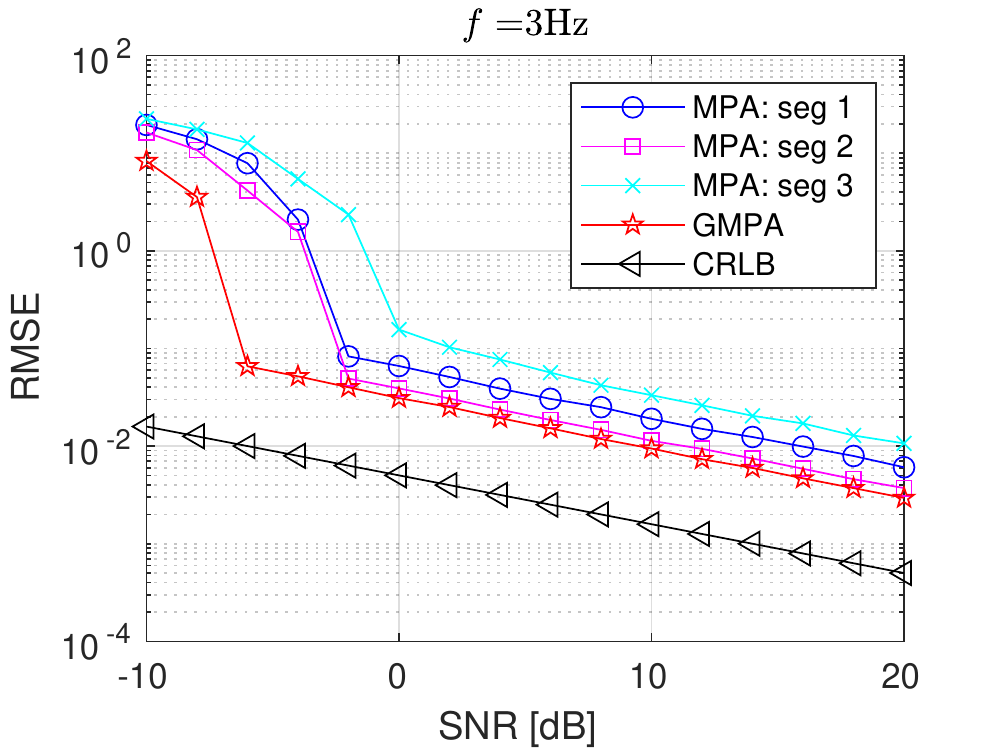}
        \label{fig:Simu_RMSE_3}
    }
    
    \vspace{-2mm}
    \subfloat[]{
    \includegraphics[width=0.36\textwidth]{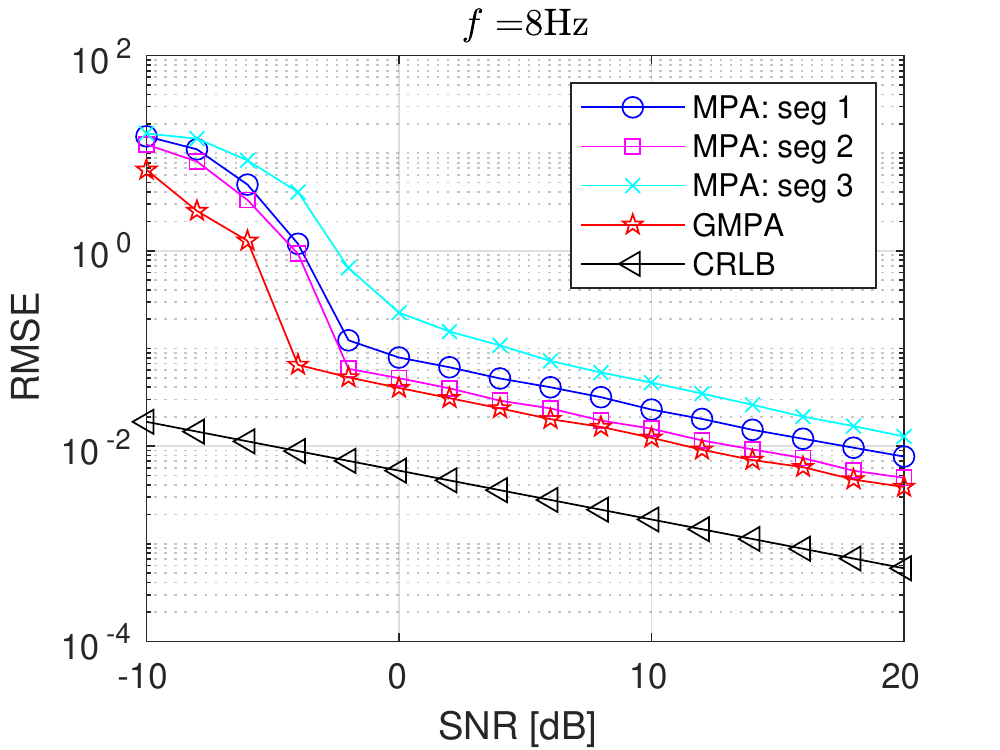}
    \label{fig:Simu_RMSE_8}
    }

    \caption{The RMSE of the estimations with each individual segment by using traditional MPA and all segments using the GMPA. The time delays between adjacent segments are integer number of sampling intervals. \protect\subref{fig:Simu_RMSE_3} and \protect\subref{fig:Simu_RMSE_8}  show the RMSEs of the estimations of $f=3\,\mathrm{Hz}$ and $f=8\,\mathrm{Hz}$, respectively. }
    \label{fig:Simu_RMSE_RTD}
\end{figure}

Moreover, the estimation accuracy of the proposed GMPA was investigated by using Monte Carlo simulations at the SNR from $-10\,\mathrm{dB}$ to $20\,\mathrm{dB}$ with steps of $2\,\mathrm{dB}$. At each SNR level, 1000 Monte Carlo runs were executed to simulate different noise implementations. The root mean square error (RMSE) of the estimations of the two frequencies with traditional MPA based on each individual signal segments and with GMPA based on all the data  are illustrated in Fig.~\ref{fig:Simu_RMSE_RTD}. The GMPA with all the data constantly obtained the most accurate estimations of both frequencies at all the SNR levels compared to traditional MPA with each individual signal segment. In particular, the GMPA with all the data obtains a significant improvement relative to the estimations based on individual segments at the low SNR (i.e., from $-4\,\mathrm{dB}$ to $-2\,\mathrm{dB}$ in this case).

\subsection{Segmented signals with arbitrary time gaps}

\begin{figure}[!t]
\centering
\vspace{-2mm}
\subfloat[]{
\includegraphics[width = 0.36\textwidth]{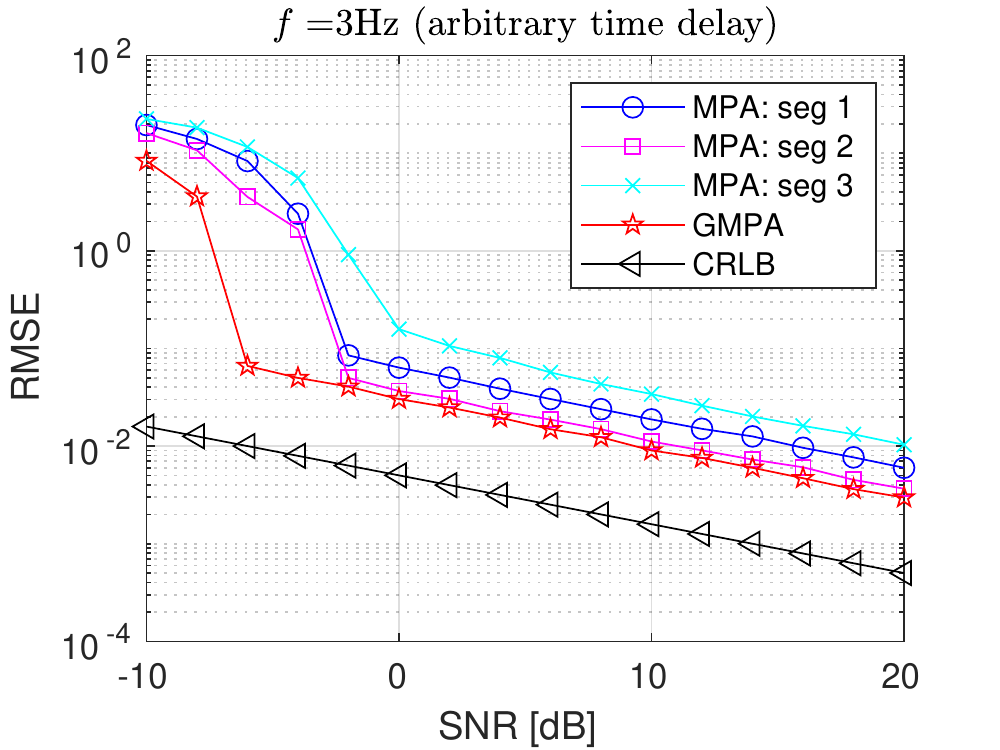}
\label{fig:Simu_RMSE_3_ATD}
}

\vspace{-2mm}
\subfloat[]{
\includegraphics[width = 0.36\textwidth]{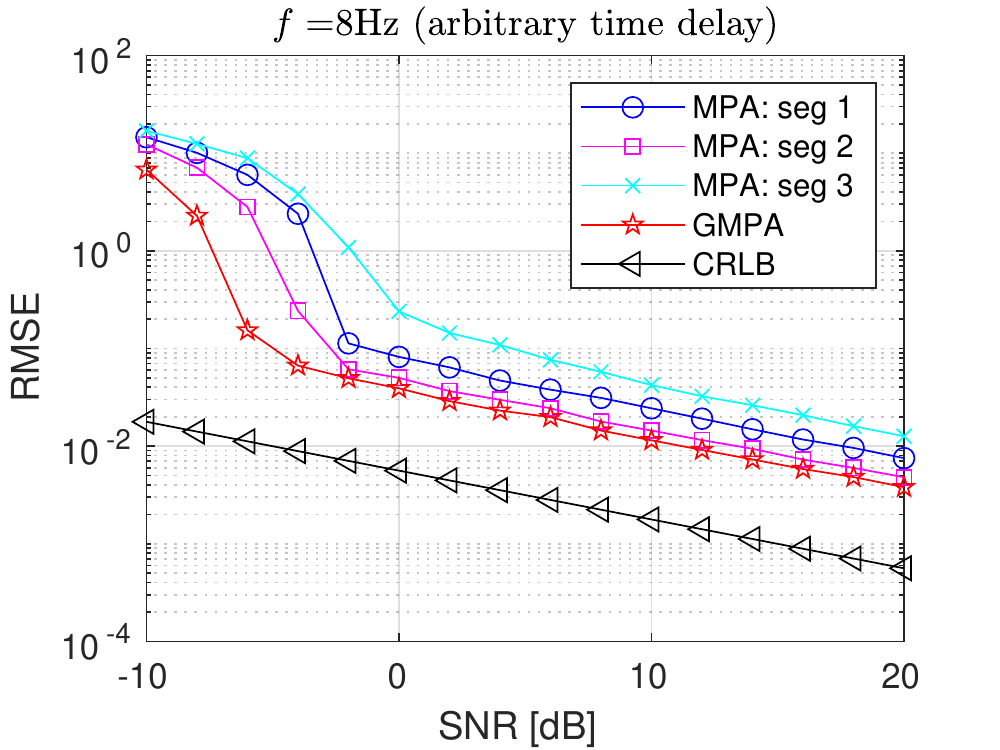}
\label{fig:Simu_RMSE_8_ATD}
}
\caption{The RMSE of the estimations with each individual segment by using traditional MPA and all segments using the GMPA. The time delays between adjacent segments are arbitrary. \protect\subref{fig:Simu_RMSE_3} and \protect\subref{fig:Simu_RMSE_8} show the RMSEs of the estimations of $f=3\,\mathrm{Hz}$ and $f=8\,\mathrm{Hz}$, respectively.}
\label{fig:Simu_RMSE_ATD}
\end{figure}

The proposed GMPA is applicable to the segmented signals acquired with arbitrary time delays between them. To demonstrate this, we kept the data in the first and third segments unchanged but modified the sampling instants $n\Delta t$ for the second signal segment as $n\Delta t + \delta t$, where $\delta t$ represents an unknown small drift from the original sampling instants. Then the time delays between the first and second segments and the second and third segments would not be an integer number of sampling intervals. As in section~\ref{subsec:seg_signal_integ_TD}, 1000 Monte Carlo simulations were run and the estimation accuracy of two frequencies with GMPA based on all the data and MPA based on each segment are presented in Fig.~\ref{fig:Simu_RMSE_ATD}. It is clear that the proposed GMPA improves the estimation accuracy by fusing the information from the signals of all three segments compared to the results obtained by the MPA with signals in each segment, especially at the low SNRs. As $\delta t$ is unknown, the proposed GMPA is, in principle, a non-coherent estimation approach; thus, its improvement of accuracy at the high SNRs is much smaller than that at the low SNRs.

{
\renewcommand{\arraystretch}{1.1}
\begin{table}[]
\centering
\caption{Estimation results of the frequencies when the segmented data are used jointly by using the GMPA method and processed separately using the MPA.}
\label{tab:Simu_Est_Freq}
\begin{tabular}{@{}lllll@{}}
\toprule
\multirow{2}{*}{}      & \multicolumn{3}{c}{MPA}                                                             & \multicolumn{1}{c}{GMPA}     \\ \cmidrule {2-5} 
                       & \multicolumn{1}{c}{seg 1} & \multicolumn{1}{c}{seg 2} & \multicolumn{1}{c}{seg 3} & \multicolumn{1}{c}{All segs} \\ \midrule
$f_1 = 3\,\mathrm{Hz}$ & 3.0191                     & 2.9949                     & 3.1234                     & 3.0139                       \\ 
$f_2 = 8\,\mathrm{Hz}$ & 8.0154                     & 8.0158                     & 7.8689                     & 7.9913                       \\ 
MSE $(\times 10^{-4})$         & 3.01                       & 1.38                       & 162.07                     & 1.34                         \\ \bottomrule
\end{tabular}
\end{table}
}

\subsection{Comparison with existing methods}

\begin{figure}[!t]
    \centering
    \includegraphics[width = 0.38\textwidth]{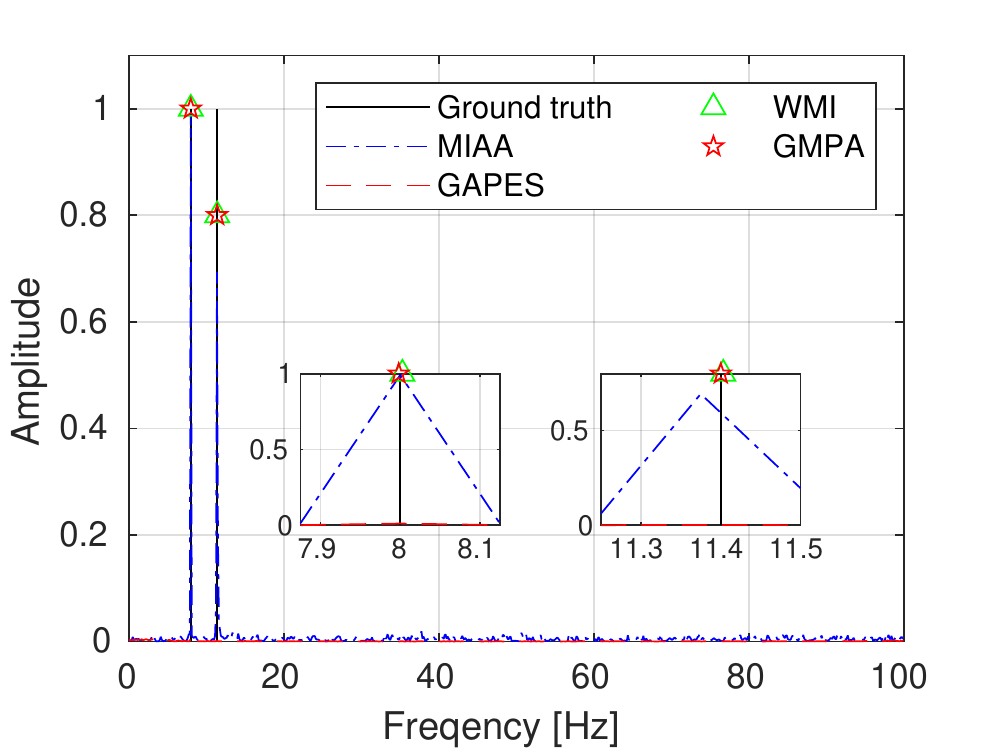}
    \caption{Comparison of estimation results of the proposed GMPA and three other state-of-the-art approaches.}
    \label{fig:simu_SOTA_comparison}
\end{figure}

In this section, the estimation accuracy of the proposed GMPA and three other state-of-the-art approaches, i.e., the MIAA \cite{Stoica2009SPL},  the GAPES~\cite{Stoica2000Astron} and the WMI \cite{Parvazi2012TSP}, are compared. A signal with two exponential components was generated with the same setup of simulation parameters in section~\ref{subsec:seg_signal_integ_TD} except that the frequencies $3\,\mathrm{Hz}$ and $8\,\mathrm{Hz}$ in the model were replaced with $8\,\mathrm{Hz}$ and $11.4\,\mathrm{Hz}$. 

After processing, the estimation results of all methods are shown in Fig.~\ref{fig:simu_SOTA_comparison}. In the implementation of MIAA and GAPES, the frequency grid with an interval of $0.125\,\mathrm{Hz}$ was used and the MUSIC criterion was used to realize the WMI approach. From Fig.~\ref{fig:simu_SOTA_comparison}, one can see that the proposed GMPA achieves the most accurate estimation, which is followed by the WMI approach. Moreover, although MIAA get accurate estimation at $f=8\,\mathrm{Hz}$, its estimation has a large estimation error at $f=11.4\,\mathrm{Hz}$, which is caused by its inherent ``off-grid'' problem that is generally suffered by Fourier basis-based approaches. By contract, the proposed GMPA directly estimate the frequencies based on the singular values of the constructed Hankel matrix; hence, it has no need to search over an predefined grid and circumvent the ``off-grid'' problem of the MIAA.

To compare the computational efficiency, all the approaches are implemented with MATLAB and run on a computer with an Intel Core i5-3470@3.2 GHz and 8GB RAM. The MIAA, GAPES, WMI and GMAP take $15.2$, $26.68$, $411.2$ and $0.0071$ seconds, respectively. Clearly, GMAP is the most efficient one.

\section{Conclusion} \label{sec:conclusion}
In this paper, a generalized matrix pencil approach (GMPA) has been presented for the estimation of exponential components with gapped data. The gaps between different data segments can be an arbitrary time interval (i.e., an interval of an integer/non-integer number of time sampling intervals). The proposed GMPA achieves the most accurate estimation compared to the existing state-of-the-art approaches and is also computationally efficient. Furthermore, it is straightforward to extend it to 2-D cases for DOA estimation with distributed arrays.

\ifCLASSOPTIONcaptionsoff
  \newpage
\fi



\bibliographystyle{IEEEtran}
\bibliography{ref}
\end{document}